# Embedding Physics Domain Knowledge into a Bayesian Network Enables Layer-by-Layer Process Innovation for Photovoltaics

Zekun Ren[1*&], Felipe Oviedo[2*&], Muang Thway[3], Siyu I.P. Tian[1], Yue Wang[1], Hansong Xue[3], Jose Dario Perea[4,2], Mariya Layurova[2], Thomas Heumueller[4,5], Erik Birgersson[6], Armin Aberle[3], Christoph J. Brabec[4,5], Rolf Stangl[3], Qianxiao Li[6], Shijing Sun[2], Fen Lin[3], Ian Marius Peters[2], Tonio Buonassisi[1, 2&]

[1]Singapore-MIT Alliance for Research and Technology SMART, 138602 Singapore
[2]Massachusetts Institute of Technology, Cambridge MA 02139, U.S.A.
[3]Solar Energy Research Institute of Singapore (SERIS), 117574 Singapore
[4]Institute of Materials for Electronics and Energy Technology (i-MEET), Friedrich-Alexander University Erlangen-Nürnberg, 91058 Erlangen, Germany
[5]Helmholtz Institute HI-ErN, Forschungszentrum Jülich, Immerwahrstrasse 2, D-91058 Erlangen, Germany
[6]National University of Singapore, 119077, Singapore
[*]These authors contributed equally.

[&] Corresponding authors:

| Zekun Ren | Felipe Oviedo | Tonio Buonassisi |
|---|---|---|
| Singapore-MIT Alliance for Research and Technology, 138602 Singapore | 77 Massachusetts Ave., Bldg. 35-135, Cambridge MA 02139 | 77 Massachusetts Ave., Bldg. 35-213, Cambridge MA 02139 |
| +65-6516-8603 | +1-617-642-1529 | +1-617-324-5130 |
| zekun@smart.mit.edu | foviedo@mit.edu | buonassisi@mit.edu |

**Abstract**: Process optimization of photovoltaic devices is a time-intensive, trial-and-error endeavor, which lacks full transparency of the underlying physics and relies on user-imposed constraints that may or may not lead to a global optimum. Herein, we demonstrate that embedding physics domain knowledge into a Bayesian network enables an optimization approach for gallium arsenide (GaAs) solar cells that identifies the root cause(s) of underperformance with layer-by-layer resolution and reveals alternative optimal process windows beyond traditional black-box optimization. Our Bayesian-network approach links a key GaAs process variable (growth temperature) to material descriptors (bulk and interface properties, *e.g.*, bulk lifetime, doping, and surface recombination) and device performance parameters (*e.g.*, cell efficiency). For this purpose, we combine a Bayesian inference framework with a neural-network surrogate device-physics model that is 100x faster than numerical solvers. With the trained surrogate model and only a small number of experimental samples, our approach reduces significantly the time-consuming intervention and characterization required by the experimentalist. As a demonstration of our method, in only 5 MOCVD depositions, we identify a superior growth temperature profile for the window, bulk and back surface field layer of a GaAs solar cell, without any secondary measurements, and demonstrate a 6.5% relative AM1.5G efficiency improvement above traditional grid search methods.

**Keywords:** solar cell, process optimization, thin-film, machine learning, Bayesian network, Autoencoder, neural network, GaAs, process innovation, deep learning

**I. Introduction:**

Process optimization is essential to reach maximum performance of novel materials and devices. This is especially relevant for photovoltaic devices, as numerous process variables can influence their performance. Often, process optimization is done using black-box optimization methods,(*e.g.*, Design of Experiments[1], Grid Search[2], Bayesian Optimization[3,4], Particle Swarm Optimization[5], etc.), in which selected variables are modified systematically within a range and the system's response surface is mapped to reach an optimum. These methods have shown potential for inverse design of materials and systems in a cost-effective manner, and are usually postulated as ideal methods for future self-driving laboratories[6-13]. However, traditional black-box optimization approaches have limitations: the maximum achievable performance improvement is limited by the designer's choice of variables and their ranges, artificially constraining the parameter space. Furthermore, insights into the root causes of underperformance are severely limited, often requiring secondary characterization methods or batches composed of combinatorial variations of the base samples. In contrast, recently, Bayesian inference coupled to a physics-based forward model and rapid, light-dependent and temperature-dependent, current-voltage measurements were shown to offer a statistically rigorous approach to identify the root cause(s) of underperformance in early-stage photovoltaic devices[14]. Furthermore, recently, the combination of physical insights with machine learning models have shown good promise in development of energy materials[15-23].

In this contribution, we consider the optimization of the synthesis temperature profile of a gallium arsenide (GaAs) solar cell using a Metal Organic Chemical Vapor Deposition (MOCVD) reactor. Growth temperature isis one of the most important and challenging parameters to optimize in III-V film deposition[24,25]. Previous studies showed that the growth temperature has an impact on the film's growth rate, surface morphology, dopant incorporation and defect formation[24,25]. Other important process parameters include precursor flowrate and growth pressure. These process parameters are closely related, and the relationship can be approximated using the Ideal Gas Law in the kinetic epitaxy process [26]. Therefore, we use the growth temperature as the key optimization variable. For other secondary process variables, *e.g.*, precursor type and carrier gas flowrate, the physical relation between process variables and materials properties is unclear and likely tool-specific[27], we can replace the physics-based parametrization in the first layer of the Bayesian network inference with a machine learning model with higher capacity, such as kernel ridge regression[28].

GaAs solar cells comprise several layers, including a back surface field, a bulk absorber, and a window layer[29].To maximize device performance, material properties need to be optimized for each layer and interface[20,30]. An experienced researcher would grow and characterize each layer (emitter, base, window and back surface field) separately to map the process variable to material properties, in an attempt to gain physical insights to optimize the final solar cells efficiency[25,31-33]. In this context, optimizing growth temperature of GaAs solar cells becomes an optimization scenario in which one process variable (temperature) affects several materials descriptors in various device layers. With the assistance of a solar cell physical simulator and additional characterization techniques, the optimal growth temperature for each layer could be pinpointed and the whole device stack could be grown using the optimized growth profile. However, this expert approach requires fabricating many auxiliary samples at varying conditions with multiple layer variations, and use secondary characterization measurements, such as Secondary Ion Mass Spectroscopy (SIMS) and Time Resolved Photoluminescence (TR-PL), to confirm root causes of underperformance. These characterization techniques are significantly more costly than current-voltage (*JV*) measurements, the primary proxy of solar cell performance. This problem mirrors the challenges in optimizing other multi-layer energy systems and semiconductors, including light-emitting diodes, power electronics, thermoelectrics, batteries and transistors.

To address this challenge, we combine several machine-learning techniques to infer the effects of a given process variable on different device layers. To avoid performing expensive characterization, such as SIMS or TR-PL, we perform automated current-voltage measurement at multiple illumination intensities (*JVi*) as the input for the algorithm. To speed up our calculations, we employ a physics-based "surrogate" model that mimics a complex physical model, in this case solar cell growth. Our surrogate model consists of a two-step Bayesian inference method, typically referred as Bayesian network or hierarchical Bayes[34-36], with relations between layers constrained by physical laws. Embedded therein is a surrogate device-physics model, which operates >100x faster (shown in Figure S1 in Supplementary Information) than a numerical device-physics solver.

Figure 1 shows the schematic of our Bayesian network. We propose three methodological innovations in this approach. First, we create a parameterized process model by imposing physics-based constraints to couple the process optimization variable (*e.g.*, growth temperature) to the resulting material's bulk and interface properties (*e.g.*, lifetime). This parametrization limits the number of fitting variables in the first layer of our Bayesian inference model, reducing the risk of overfitting, and provides a degree of interpretability. Second, we add another inference layer inside a numerical device-physics model, linking the inferred bulk and interface properties to the solar cell performance measures (*e.g.*, *JVi* characteristics, quantum efficiency, and energy conversion efficiency). Extraction of underlying materials descriptors from *JVi* curves, previously demonstrated in Ref[14], enables us to trace the root cause(s) of device underperformance to a specific material or interface property. Third, we achieve a >100x acceleration in inference by replacing the solar cell model, a traditional PDE (Partial Differential Equation) numerical model, with a highly-accurate neural network surrogate model. This enables us to update the posterior probability distribution for our Bayesian network inference model over a vast parameter space.

In Figure 1, we also show the difference between our Bayesian-network-based optimization and the traditional black-box optimization. As only low cost evaluations (*JVi* measurement) are performed for solar cell characterization, accurate extraction of underlying material properties requires performing Bayesian inference using a device-physics model[14]. Traditional optimization approaches often make use of a purely black-box surrogate model[3] to map the relation between process variables and device performance directly, without any insights about material properties in the device. In contrast, our Bayesian network inference connects target variables to material descriptors, then to process conditions. It provides rich, layer-by-layer information about critical material properties that affects device's electrical performance. In this study, we chose to map doping concentration in emitter and bulk, bulk lifetime ($\tau$) , front and rear (Indium Gallium Phosphide) InGaP/GaAs surface recombination velocities (*SRVs*) to growth temperature using and customize the growth temperature that maximize those desired material properties Replacing the traditional optimization (process variable-device performance) with our Bayesian-network-based optimization (process variable-material properties-device performance) feasibly enables us to expand the variable space, and identify design process windows that selectively improve specific materials, layers and interfaces inside a solar cell. This results in vastly improved device performance and process interpretability in few MOCVD fabrication rounds with a single temperature sweep.

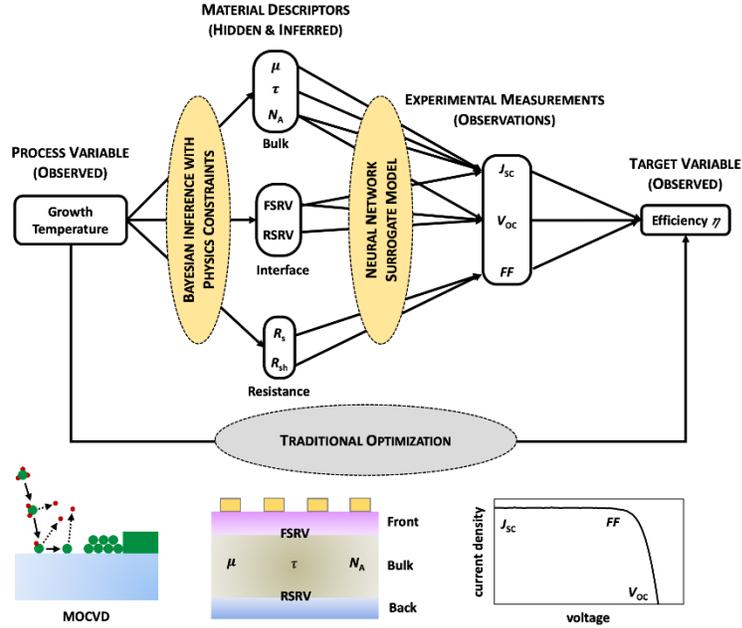

*Figure 1. Schematic of our Bayesian-network-based process-optimization model, featuring a two-step Bayesian inference that first links process conditions to materials descriptors, then the latter to device performance. Our Bayesian-network-based process-optimization back propagates from efficiency to bulk interface properties then to growth temperature, enabling layer-by-layer tuning of process variables.*

To demonstrate the potential of our approach, we use our Bayesian network to characterize and optimize, in a single temperature sweep consisting of 5 MOCVD fabrication rounds, the process temperature of a GaAs solar cell. Our devices have a baseline efficiency of ~16% without an Anti-Reflection Coating (ARC). Our Bayesian network approach identifies the optimal set of process conditions that translate into maximum performance under the physical model and real process constraints. The physical insights from the Bayesian network inference suggest an optimal growth temperature profile, allowing a significant 6.5% relative increase in average AM1.5G efficiency above baseline in a single temperature sweep (6th MOCVD run).. This result verifies the capacity of our approach to find optimal process windows with little intervention from the experimentalist, no secondary characterization techniques or auxiliary samples, and with performance beyond experimentalist-constrained optimization.

**II. Results and Discussion:**

Bayesian network

As illustrated in Figure 1, we construct a Bayesian network to link the process variables with each material and device property in the GaAs solar cells. Hereafter, we optimize each material property separately to maximize the final device performance. The Bayesian network consists of four parts:

*1. Parameterization of process variables by embedding physics knowledge:*

This section describes how we define physics-based relations between process variables and materials descriptors, embedding physics domain knowledge, and ensuring faster and better convergence of our Bayesian optimization algorithm. This corresponds to the progression from "Process Conditions" to "Materials Descriptors" in Figure 1. Device fabrication of solar cells is expensive, thus it is essential to explore the process variable space efficiently[37]. From a machine-learning point of view, we leverage the existing knowledge from literature and embed such domain knowledge as prior parameterization to constrain the variable space, *e.g.*, Equation [2]. The parameterization connects process variables with underlying material and interface properties. In this study, we chose to infer emitter and bulk doping concentration, bulk lifetime ($\tau$), front and back surface recombination velocity (*SRV*) as the intermediate material properties because they play an critical role in determining the device' electrical performance[29] and each property is layer or interface specific. In the case of chemical vapor deposition (CVD), recognizing that growth temperature affects several thermally and kinetic activated processes[38], we embed such knowledge and enforce an exponential dependence of underlying material properties based on the modified Arrhenius equation[39-41] (Equation [2]).

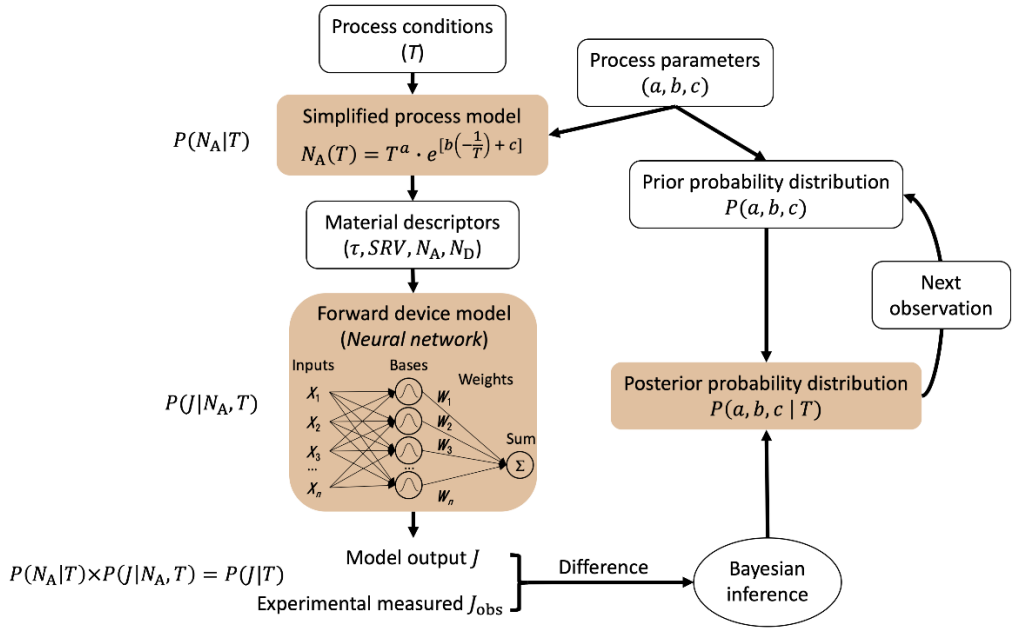

*Figure 2. Architecture of our Bayesian inference network, to identify new windows for process optimization.*

The detailed schematic of the Bayesian network inference is shown in Figure 2. To illustrate the flow of our approach, we use the optimization of acceptor (Zn) doping concentration in a GaAs solar cell as a showcase. Our approach can be represented as a two-step Bayesian inference procedure using conditional probability (Equation [1]).

$$P(J|T) = P(N_A|T) * P(J|N_A,T) \qquad [1]$$

P ($N_A|T$) is the conditional probability of Zn acceptor doping levels given the process temperature. We parameterize the prior (P ($N_A|T$)) based on existing literature and our physical knowledge. Recognizing that MOCVD growth is a kinetic process[38], we enforce an Arrhenius equation-type of parameterization to link the underlying material properties with growth temperatures. Zn doping level can be represented in the modified Arrhenius Equation [2].

$$N_A(T) = T^a \cdot \exp\left[ b\left(-\frac{1}{T}\right) + c \right] \qquad [2]$$

(*a, b, c*) are latent process parameters that are inferred from the Bayesian framework. *b* and *c* correspond to the activation energy and pre exponential factor in the traditional Arrhenius equation. *a* is the temperature dependence of the pre-exponential factor (ln(*c*)). Aside from Zn doping concentration, Si doping concentration, bulk minority carrier lifetime ($\tau$), and front and back *SRV*'s are also parameterized in the same fashion.

The modified Arrhenius equation form for the doping concentration agrees well with trends reported in literature[42-44]. There is insufficient literature and domain expertise to directly relate bulk and interface properties with the growth temperature. However, previous study has shown that $\tau$ and *SRV*'s are correlated with doping concentration[45,46]. Note that performing the fitting of Equation [2] can be an implicit hypothesis test. A small *a* value suggests that the pre-exponential factor temperature dependence is suppressed, and that the Arrhenius relationship governs the temperature dependence of the particular bulk, interface, or resistance property. On the other hand, a big *a* value suggests a larger contribution of the pre-exponential factor to temperature dependence, indicating a deviation from a pure

Arrhenius-like regime at a given temperature. Additional domain knowledge is embedded in the prior by setting hard constraints for the material properties. The ranges of the five inferred material properties are shown in Table S2.

*2. Inference of material and device properties from device measurements*

This section describes the progression from "Materials Descriptors" to "Target Variable: Performance" in Figure 1. Inference of underlying material properties from *JVi* measurements is used to trace the root cause(s) of device underperformance to specific material or interface properties. We further extend the connection between process variables and materials properties to device measurements by adding an additional inference layer. The forward model of this inference layer is a numerical device-physics model, linking the inferred bulk and interface properties to solar cell device parameters (*e.g.*, current-voltage characteristics, quantum efficiency, and conversion efficiency).

Following the above example, $P(J|N_A,T)$ is the conditional probability of a set of *JVi* observations at a series of fixed illumination intensities given the underlying material parameters (Zn doping concentration). The material property-*JVi* relation is extensively investigated and can be solved numerically using a well-developed device model from literature[29,47-49]. A well-calibrated PC1D model[47] is used in this work. However, numerical simulation is computationally expensive in the Bayesian framework (which requires hundreds of thousands of simulated runs to provide adequate posterior probability estimation) and makes it difficult to integrate new features into the model. Furthermore, experimental *JVi* observations contain experimental noise that causes deviations from simulated *JVi* curves.

*3. Replacement of numerical solver with a robust neural-network surrogate model*

To circumvent the computational complexity of the numerical device-physics model and the discrepancies between experimental and simulated *JVi* curves, we replace the numerical simulator with a surrogate deep neural network. Figure 3 shows a schematic of the model, consisting of two parts: (1) A de-noising Autoencoder (AE) [50] that takes noisy *JVi* curves as input and reconstructs noise-free *JVi* curves. In our case, the training data are 20,000 simulated *JVi* curves, computed with a device-physics model, and augmented with Gaussian noise that mimics experimental noise. The Gaussian noise has a 0 mean and 0.2% variance which are estimated from the repeated *JV* measurements. The structure of the encoder network is shown in Figure S2, and consists of 3 convolutional and 2 dense layers in the encoder and 3 convolution transpose and 2 dense layers in the decoder. The decoder is a mirror of the encoder network, with transposed convolutions layers replacing the convolutional layers. The de-noising training of the architecture provides robustness to experimental noise. (2) A regression model that predicts the *JVi* curves based on underlying material properties. The regression model has the same structure as the decoder used in the de-noising AE. $P(J|N_A,T)$ thus can be computed using this surrogate neural-network model.

To create the training dataset, we first randomly sample a set of 20,000 random material properties ($\tau$, *FSRV*, *RSRV*, Zn and Si doping concentration) from uniform probability distributions. The threshold of the uniform distribution is shown in Table SI. Then, we use scripted-PC1D[51] to numerically simulate a set of 20,000 *JVi* curves from the chosen material descriptors. Although domain expertise is required in setting up the numerical PC1D model, this exercise is a one-time implementation for each material system. Subsequently, we augment the simulated *JVi* curves with Gaussian noise to mimic the experimental measurements and feed the noisy *JVi* to train the Denoising AE. Figure 3 shows the noise-free *JVi* curves after we feed the experimental data to the AE. Hereafter, we train the neural network regression model to predict *JVi* curves from material descriptors in the latent space of the AE. The loss for both AE and regression model is chosen to constraint the latent parameter space to the 5 variables of interest, and is minimized using the ADAM gradient descent algorithm with a batch size of 128 and an initial learning rate of 0.0001. We split the *JVi* curves into 80% and 20% for training and testing purposes. The numerical solver in the Bayesian network is then replaced by the regression model. The surrogate model is significantly more computationally efficient than the numerical solver. Figure S1 shows the acceleration by adapting the neural network surrogate for calculation of a set of *JVi* curves. The surrogate model, running on a GPU, is 130 times faster than the PC1D numerical solver and 700 times faster if the numerical solver is called externally.

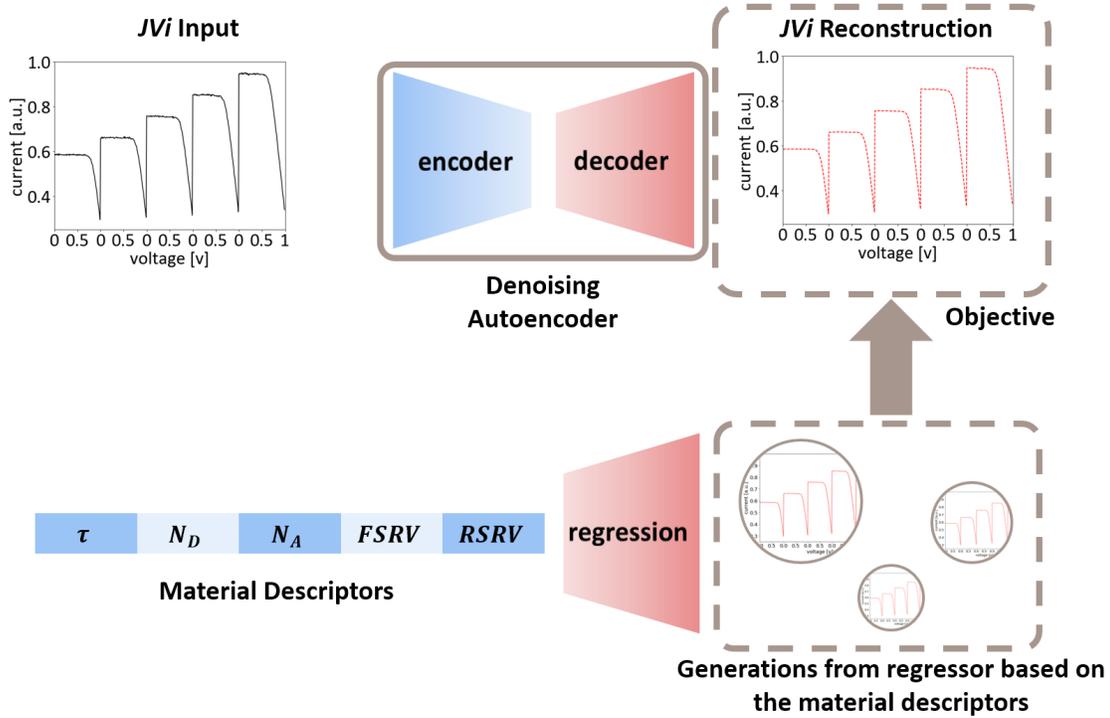

*Figure 3. Schematic of neural network surrogate model to infer material descriptors from JVi curves. (In this figure, 5 sequential JV curves are shown as inputs and outputs, with increasing illumination intensity.)*

Once the device model is trained, we connect these previous two Bayesian inference steps into a hierarchical structure using Equation [1]. A posterior probability to every combination of latent fitting parameters (*a, b, c*) is assigned. This probability is represented by a multivariate probability distribution over all possible combinations of model fit parameters. This probability is modified every time new data (*JVi* measurement) is observed. We apply a Markov Chain Monte Carlo (MCMC) method for sampling the posterior distribution of latent parameters (*a, b, c*); this achieves an acceleration of Bayesian inference computation time comparable or superior to the successive grid subdivision method [2]. Specifically, we implement the affine-invariant ensemble sampler of MCMC proposed by Goodman & Weare[52]. With each newly observed *JVi* measurement, the posterior distributions of the latent process parameters (*a, b, c*) are updated. Hereafter, the material descriptor (Zn doping concentration as a function of growth temperature ($N_A(T)$) can be obtained from Equation [2].

In an analogous way, other descriptors, such as the doping levels of other species and bulk and interface recombination properties, can be obtained as a function of the process variables and adequate prior parametrizations. We use this result to optimize the MOCVD growth temperature of a set of GaAs solar cells.

*4. Optimizing solar cells using our Bayesian network inferred results:*

After we map the growth temperature to the material properties, we apply grid search method with 10°C resolution to find the growth temperature that maximizes the desired material properties and minimize the undesired properties (maximize $\tau$ and minimize *SRVs*) for the solar cell . Mathematically, we can define the optimization procedure enabled by our Bayesian network model as:

$$x^* = \arg\max(h(g_i(x))) \qquad [3]$$

The variable x* is the set of process variable, specifically the MOCVD growth temperature, that produce the largest solar cell efficiency. We first estimate the function $g_i(x)$, which models how the set of underlying material properties changes with the process variable. Hereafter, the cell efficiency can be maximized by plugging material properties $y_i = g_i(x)$ to $h(y_i)$, which models the relation between material properties and the final solar cell performance (*JVi* curves). $h(y_i)$ can be solved numerically and, in our case, is estimated using a neural network. The material properties extracted can be exploited to find $x^*$ that maximizes the cell efficiency. As $g_i(x)$ determines the functional relation of materials descriptors and the process variable, we can tailor our process variable to maximize the desired materials properties, such as lifetime, and minimize the undesired properties, such as *SRVs*, in selected locations across the devices.

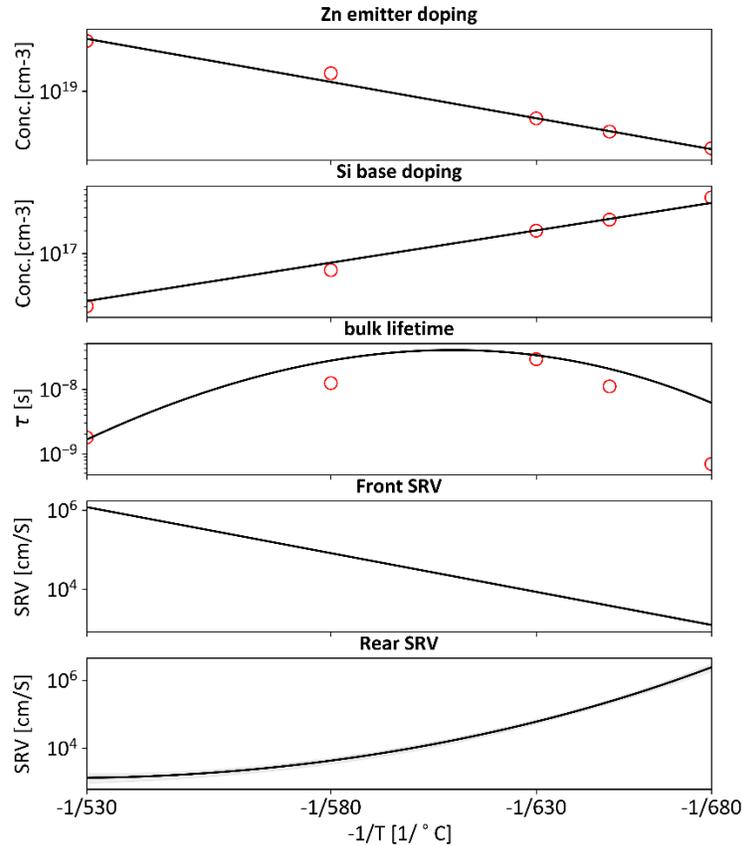

*Figure 4. Bayesian network reveals how each material descriptor (bulk and interface property) varies with processing conditions. Black lines show inferred values of materials descriptors as a function of growth temperature; red circles show experimental validation of materials descriptors using SIMS and TR-PL. Doping concentrations of different species (Zn and Si), bulk lifetime, and InGaP/GaAs interface SRV can be inferred from finished device measurements using this approach. The x-axis is -1/T and the y-axis is in logarithm scale to illustrate whether the material property follows the Arrhenius equation (linear trend).*

Table I Mean value of latent parameter (*a,b,c*) for different material properties

|  | *a* | *b* | *c* |
|---|---|---|---|
| Zn doping | 0.0018 | -0.1494 | -0.1948 |
| Si doping | 0.0016 | 0.1551 | 0.2970 |
| Bulk lifetime | -4.5973 | 2.7984 | 2.3687 |
| Front SRV | 0.0015 | -0.1440 | -0.1892 |
| Rear SRV | 2.1194 | -1.1119 | -0.7300 |

As a baseline for testing our methodology, we fabricate 5 batch of GaAs solar cells (4 cells per batch), sweeping a range of constant growth temperatures. The GaAs solar cell structure consists of multiple InGaP and GaAs layers (figure S3), and all solar cell layers are grown at the same temperature in one MOCVD experiment. In 5 experiments, a temperature range of 530°C to 680°C, with 20–50°C temperature intervals, is explored. The films are fabricated into 1 cm$^2$ solar cells, without ARC's. Detailed growth and fabrication procedures can be found in the Experimental Procedures section. *JVi* measurements under multiple illumination intensities (0.1–1.1 suns) are performed. Figure 4 shows the inferred material properties as a function of MOCVD growth temperatures. We can see that the logarithm value of *p*-type (Zn) doping level, *n*-type (Si) doping level and *FSRV* have an almost linear correlation with 1/T, suggesting a good agreement with the standard Arrhenius Equation while the bulk lifetime and RSRV exhibits nonlinear relationships. To trace the root causes, mean of (*a, b, c*) values for each material properties after the MCMC run are attached in Table I. The full distribution of the (*a, b, c*) values extracted from the Bayesian network is shown in Figure S4. The *a* values for both the Zn and Si doping concentration are close to zero (<0.002) indicating a negligible temperature dependence in the pre-exponential factor and the Arrhenius regime is dominant. This agrees well with the trend reported in literature for various dopant species [42-44]. The *a* value for *FSRV* is also insignificant, and *FSRV* has similar trend as the Zn doping concentration. We postulate that this is due to *SRV*'s being affected by doping concentrations[45,46], and the dominant recombination mechanism in the front interface being related to Zn doping level. The *a* value for effective bulk lifetime is significant (-4.59) indicating a strong temperature dependence on the pre-exponential factor and thus non-Arrhenius regime. We postulate that this could result from the existence of both Zn and Si dopant in the GaAs bulk layer, as there is a competing contribution from the two dopants. The *a* value for *RSRV* lies between the value of bulk lifetime and Si doping levels. The *RSRV* slightly follows the linear trend of Si doping levels, however, we postulate that the subsequent bulk, front and contact layers' growth impact on the rear interface's quality[45], and contributes to the non-Arrhenius behavior.

To validate the inferred doping concentrations and lifetime from our Bayesian network approach, we grow auxiliary structures (*e.g.* single layer structure to conduct SIMS measurement and heterostructure for TR-PL measurements). The red circles in the first three subplots of Figure 4 represent the results from those independent auxiliary measurements. The experimental details are shown in the supplementary information (Figure S5). It is evident that the independent measurements agree well with the inferred material properties.

It is interesting to note that each parasitic recombination parameter (bulk lifetime, *FSRV* and *RSRV*) has its minimum/maximum at a different growth temperature. The bulk lifetime peaks around 620°C, which is close to our baseline process temperature (630°C). The front and rear *SRVs* exhibit opposite trends when growth temperature increases. Instead of growing the whole GaAs stack at the same temperature, Figure 4 indicates that the back contact, bulk, and front contact should each be grown at a different temperature to optimize performance.

Table II. Temperature profile of GaAs solar cells

| Layer | Optimal growth temperature suggested by Bayesian Network [°C] |
|---|---|
| Buffer | 580 |
| InGaP BSF | 580 |
| GaAs bulk | 620 |
| InGaP window | 650 |
| GaAs contact | 650 |

This knowledge gained by the Bayesian network enables us to formulate a new time-temperature profile (Table II) for our GaAs devices (labeled "Bayes Net" in Figure 5). We performed an additional MOCVD experiment by selecting the growth temperature show in Table II that minimizes recombination at each layer or interface in a 10°C resolution (hardware tolerance) grid. We avoid extreme conditions (*e.g.*, 680°C), which show deterioration of overall device performance (Figure 5) despite inferred layer-specific improvements (Figure 4).

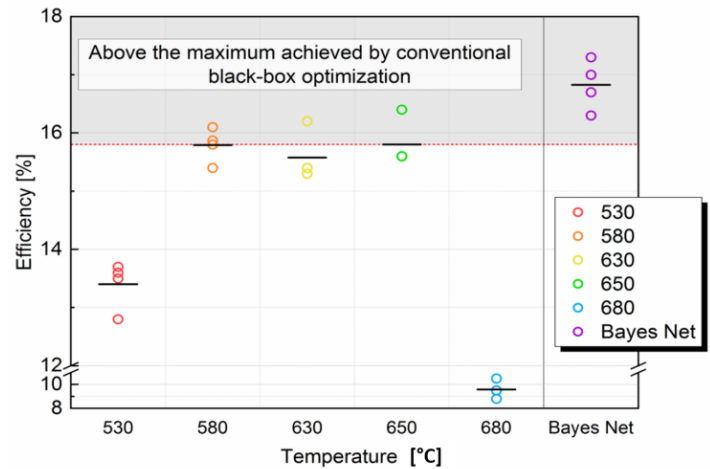

*Figure 5. Comparison of "grid search optimization" versus our approach using a Bayesian network (Bayes Net). GaAs cell efficiency varies as a funciton of growth temperature, reaching an average maximum between 580 and 650°C. Our Bayesian-network-informed process (labeled "Bayes Net") tunes the growth temperature of each layer to minimize recombination (Figure 4), increasing efficiency by 6.5% relative. The grey area represents the additional efficiency gain that cannot be achieved using conventional grid search optimization. Please, note that black-box optimization methods searching in the constant temperature space would have underperformed compared to the Bayesian network results.*

Figure 5 shows the spread of GaAs cells efficiency for the 5 MOCVD experiments and the additional MOCVD run with the customized growth profile. Without additional insights on materials properties from our Bayesian network, cell efficiency become the sole optimization target. The grid search on growth temperature suggests growing the whole solar cell stack at 580°C or 650°C are the optimal growth scheme. This temperature sweep (*i.e.* a single cycle of learning) gives us an efficiency improvement of 1.4% relative above our baseline efficiency (630°C) after 5 MOCVD runs. The 6$^{th}$ MOCVD run that tunes growth temperatures of each layer (Table I), thereby minimizing layer-specific recombination achieve a 6.5% relative improvement over the baseline, well exceeding the conventional approach. Auxiliary one-Sun *JV* and external quantum efficiency (*EQE*) measurements are performed to trace the root-causes of efficiency improvement using the customized temperature profile (Figure 6). It shows that both $J_{SC}$ and $V_{OC}$ are responsible for the efficiency improvements in our "Bayes Net" growth temperature profile. *EQE* shows that photo-response at wavelengths less than 820 nm (corresponding to an optical penetration depth comparable to our 2-µm thick absorber) is improved, indicating significant reduction in recombination for the front and bulk layers. We perform Bayesian inference (second layer in the Bayesian network) to extract the material properties of measured *JVi* curve of this cell and our baseline. The mean values of *FSRV*, *RSRV*, and $\tau$ of the cells grown using "Bayes Net" profile (Table II) are $1.2\times10^3$ cm/s, $5.4\times10^4$ cm/s, and 29 ns, while the best baseline values are $4.1\times10^3$ cm/s, $6.1\times10^4$ cm/s, and 26 ns. These values agree qualitatively well with the EQE observations from auxiliary measurement, which show the front surface and the bulk benefiting the most from the "Bayes Net" temperature profile, possibly because these were the highest-temperature steps, and that may have partially erased the thermal history of the underlying rear-surface layer.

All cells reported herein do not have anti-reflection coatings; the best cells shown in the figure are estimated to have efficiencies in the 24% to 25% range with optimal double-layer ARC's. The efficiency value is near state-of-the-art for a single-junction GaAs with substrate[53] grown in an academic setting. Other process parameters, *e.g.*, epitaxial lift-off and contact grid optimization, are required to reach record efficiencies[30]. Nevertheless, the recombination gains enabled by the variable-temperature profile by our Bayesian network should translate to these advanced cell architectures. It is important to note that, given the shape of the function to optimize, any other black-box optimization methods in the constant temperature space would have underperformed in comparison to the Bayesian network. Growing the device stack at the constant temperature will never achieve the same level of improvement as what is demonstrated using the Bayesian network. This is the case because tuning layer-by-layer growth temperature only becomes evident when we perform Bayesian inference to map the *JVi* measurements to underlying material properties. This demonstrates how additional insights gained via Bayesian-network-based optimization can be translated into device performance that exceeds black-box optimization. One could argue that similar performance can be achieved by following the "expert" approach to perform single-layer optimization before incorporating them into a device stack. However, many auxiliary structures growth and secondary measurements will be required in this case. 15 SIMS and TR-PL samples were grown in this study for model validations. The fact that the optimal variable temperature profile is found after a single temperature sweep of 5 MOCVD runs at constant temperatures, verifies the potential of our method to accelerate time-consuming device optimization significantly, limiting the need of synthesizing auxiliary samples and performing secondary measurements. Lastly, we can modify the Bayesian Network approach by replacing the physics parametrization (Arrhenius equation) with regularized black-box regression, in those cases when the physics between the process variable and material properties are unclear or complex to model. In the Figure S6 supplementary, we demonstrate that the temperature can be mapped to the material properties manifold with a similar accuracy by replacing the Arrhenius equation parametrization with a black-box regression model (kernel ridge regression using radial basis function[28]). Incorporating black-box regression in the first layer of the Bayesian network enables one to describe complex process variables. However, the performance of the black-box regression will be affected by the hyperparameter values shown in Figure S6. Because the experiments are expensive in our case and data scarcity, the initial selection of hyperparameters in the black-box regression can be critical. Furthermore, the interpretability of the Bayesian network will surfer as latent process parameters (*a, b, c*) cannot be inferred in the black-box regression case.

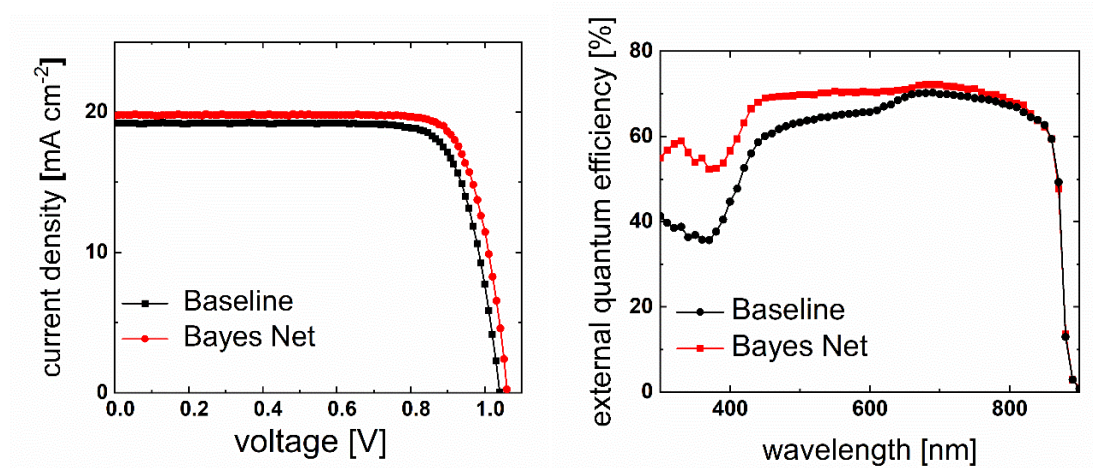

*Figure 6. JV and EQE measurement of GaAs solar cells with the custom growth-temperature profile informed by our Bayesian network (red) and baseline 630°C (black)*

**III. Conclusions:**

We developed and applied a Bayesian-network approach to GaAs solar cell growth optimization. This approach enables us to exceed our baseline efficiency by 6.5% relative, by tuning process variables layer-by-layer, in just 6 MOCVD experiments. Our approach is enabled by implementing physics-informed relations between process variables and materials descriptors, and embedding these into a Bayesian network. We link these materials descriptors to device performance using a neural network surrogate model, which is 100x faster than numerical device simulation. The small number of growth (MOCVD) runs necessary to implement this layer-by-layer process-optimization scheme translate into significant cost and time reductions compared to common black-box optimization methods. We believe this approach can generalize to other solar cell materials[54,55], as well as other systems with physics-based or black-box relations between process variables and materials descriptors, and physics-based device-performance models. Our surrogate model can replace common models in closed-loop black-box optimization, such as a Gaussian Process regression in Bayesian optimization, providing good functional fitting and physical insights.

**IV. Method**

Experimental Procedures: The top GaAs cell was fabricated on epi-ready <100> oriented GaAs on-axis wafers using an AIXTRON Crius MOCVD reactor. The growth is performed under a reactor pressor of 100mbar using TMGa, TMIn, $AsH_3$ and $PH_3$ as precursors and 32 standard liters per minute (slm) $H_2$ as carrier gas. It has a 3 µm *n*-doped GaAs base (Si dopant) and 100 nm $p^+$-doped GaAs emitter (Zn dopant). Highly doped InGaP is used as the window (Zn dopant) and back surface field (BSF) layer (Si dopant). $p^+$-doped GaAs layer (carbon dopant) is added at the front surface to achieve an ohmic contact to the metal fingers. The solar cells are metalized using an E-beam evaporator and a shadow mask to fabricate a grid pattern with ~8% shading. A SIMS measurement is conducted for the GaAs films that are grown in the same batch before metallization. We additionally grow $n$-doped InGaP/GaAs/InGaP heterostructure with two different base thicknesses (500nm and 1000nm) to measure the bulk lifetime of the $n$-doped GaAs bulk[56]. The growth conditions for the heterostructure are identical to the conditions for GaAs solar cells.

Data availability: The experimental and simulated dataset for training the neural network surrogate model and predicting the optimal growth conditions from Bayesian network is available in the following GitHub repository: https://github.com/PV-Lab/BayesProcess. Additional data supporting the findings of this study is available from the authors upon reasonable request.

Code availability: The code used for Bayesian network and neural network surrogate to predict material properties is also available at https://github.com/PV-Lab/BayesProcess.

**Acknowledgements:** We acknowledge Kaicheng Zhang and Ning Li from FAU. We acknowledge Chuanseng Tan from NTU for providing fabrication facilities for the cell development at Nanyang NanoFabrication Centre. This research is supported by the National Research Foundation, Prime Minister's Office, Singapore under its Campus for Research Excellence and Technological Enterprise (CREATE) programme and its Energy Innovation Research programme EIRP-13 (Award No. NRF2015EWT-EIRP003-004) (supporting GaAs device fabrication), by the National Research Foundation (NRF) Singapore through the Singapore Massachusetts Institute of Technology (MIT) Alliance for Research and Technology's Low Energy Electronic Systems research program (supporting autoencoder and physics-constrained Bayesian inference algorithm development), by the U.S. Department of Energy Photovoltaic Research and Development Program under Award DE-EE0007535 (supporting Bayesian optimization algorithm development), and by a TOTAL SA research grant funded through MITei (supporting ML algorithm framing and application).

**Author Contributions:** Z.R. and F.O. contributed equally for this work. Z.R., F.O., and T.B. conceived of the research. Z.R. and F.O. developed and tested the Bayesian network, with key intellectual contributions from H.X., J.D.P., M.L., T.H., C.J.B., R.S., S.S., I.M.P., and T.B. Z.R. and F.O. developed the neural network surrogate model for fast inference. Z.R. fabricated the GaAs solar cells with the help from Y.W., M.T., A.A., and F.L. Z.R., F.O, and T.B. wrote the manuscript, with input from all co-authors.

**Declaration of Interest:** The authors declare no competing interests.

# Embedding Physics Domain Knowledge into a Bayesian Network Enables Layer-by-Layer Process Innovation for Photovoltaics

Zekun Ren[1*&], Felipe Oviedo[2*&], Muang Thway[3], Siyu I.P. Tian[1], Yue Wang[1], Hansong Xue[3], Jose Dario Perea[4,2], Mariya Layurova[2], Thomas Heumueller[4,5], Erik Birgersson[6], Armin Aberle[3], Christoph J. Brabec[4,5], Rolf Stangl[3], Qianxiao Li[6], Shijing Sun[2], Fen Lin[3], Ian Marius Peters[2], Tonio Buonassisi[1, 2&]

[1]Singapore-MIT Alliance for Research and Technology SMART, 138602 Singapore
[2]Massachusetts Institute of Technology, Cambridge MA 02139, U.S.A.
[3]Solar Energy Research Institute of Singapore (SERIS), 117574 Singapore
[4]Institute of Materials for Electronics and Energy Technology (i-MEET), Friedrich-Alexander University Erlangen-Nürnberg, 91058 Erlangen, Germany
[5]Helmholtz Institute HI-ErN, Forschungszentrum Jülich, Immerwahrstrasse 2, D-91058 Erlangen, Germany
[6]National University of Singapore, 119077, Singapore
[*]These authors contributed equally.

# I. Computation time comparison between the neural network and numerical solver:

The simulated *JVi* curves consists of 30 *JV* curves (5 different process conditions with 6 illumination intensities). The Bayesian network has MCMC chain with 20,000 samples indicating that those 30 *JV* curves are computed 20,000 times to update the posterior. Using the numerical solver, in this case, is extremely time consuming. Herein, we compare the runtime of simulating 30 *JV* curves using the numerical solver and the neural network surrogate model as shown in Figure S1. PC1d is chosen in our work as it is one of the fastest and well-studied numerical solver[1]. Although other Python-based numerical solver has been developed[2], its computation speed is not faster. In Figure S1, runtime of the full *.exe* run accounts for calling the scripted PC1d[3] program in Python. To eliminate the external file reading time from the numerical solver to Python, we also generate 30 *JV* curves using the batch mode within the PC1d program. Figure S1 shows that generating 30 *JV* curves using the autoencoder with a GPU is more than 130 times faster than generating it within the PC1d program and 700 times faster if calling PC1d externally in Python. The GPU used in this work is a Nvidia GTX 1070 and the CPU used is an Intel i7 3770.

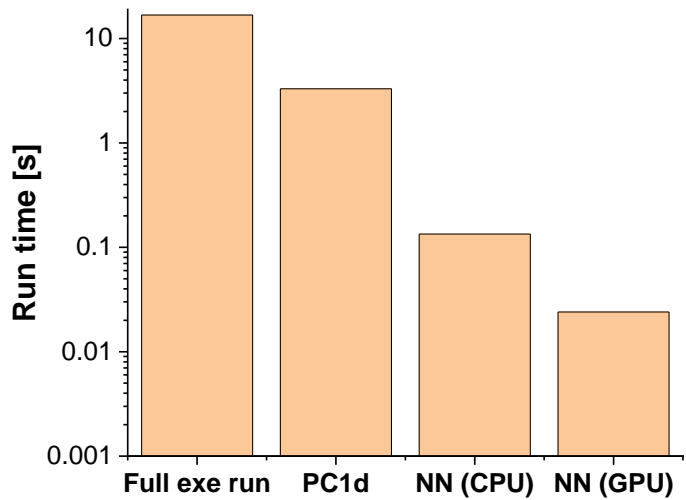

Figure S1. Computation time of 30 JV curves that are updated every MCMC run using neural network (NN) and Numerical solver (PC1d). Computation time of the full .exe run accounts for calling the scripted PC1d program in python. Computation time of PC1d only accounts for numerically generating JVs within the program.

## II. Schematic of encoder used in the autoencoder surrogate model:

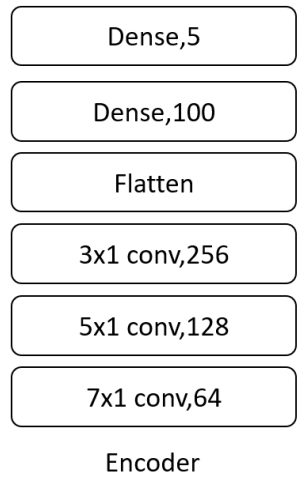

*Figure S2. a) geometry of the encoder, the decoder is a mirror of the encoder with convolutional transpose layers.*

The geometry of the encoder neural network is shown in Figure S2. (2) the decoder that takes the latent parameters and generates the *JVi* curves. The decoder is a mirror of the encoder network, with transposed convolutions layers replacing the convolutional layers.

## III. GaAs solar cell architecture:

The following schematic (not to scale) shows the GaAs solar cell fabricated according to the Experimental Procedures section.

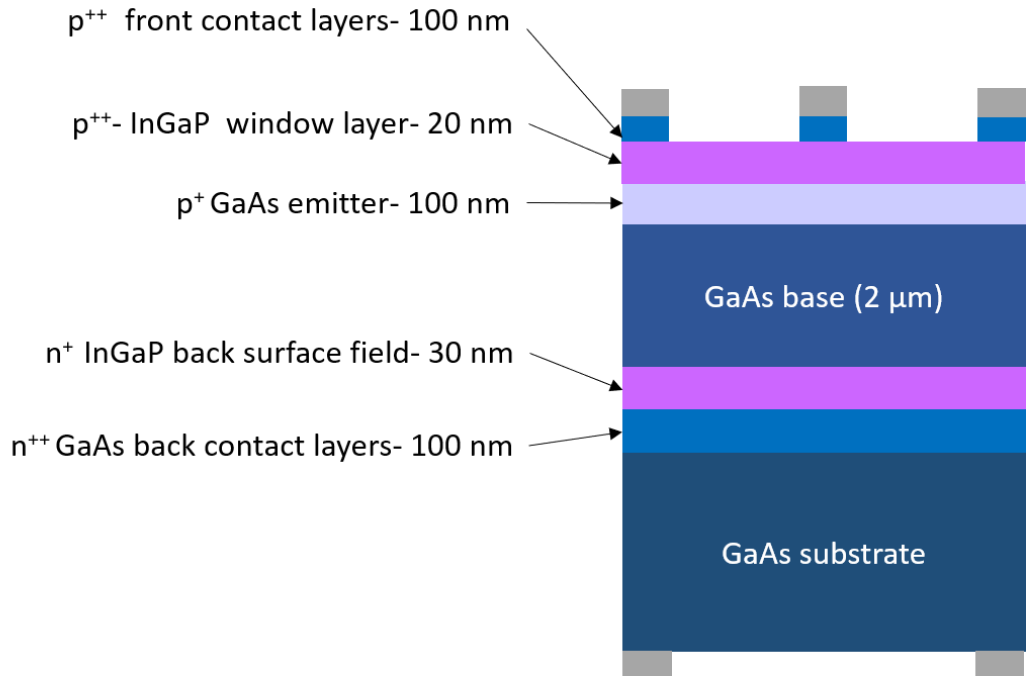

*Figure S3. Schematic (not to scale) of the GaAs solar cell architecture used in this study.*

## IV. Full distribution of latent parameters from Bayesian network:

Figure S4 is the corner plot of 15 latent process parameters ($(a,b,c)$ for 5 different materials properties) after the Markov Chain Monte Carlo(MCMC) run. The MCMC chain is conducted using an affine-invariant ensemble sampler of MCMC [4] with parallel tampering[5].

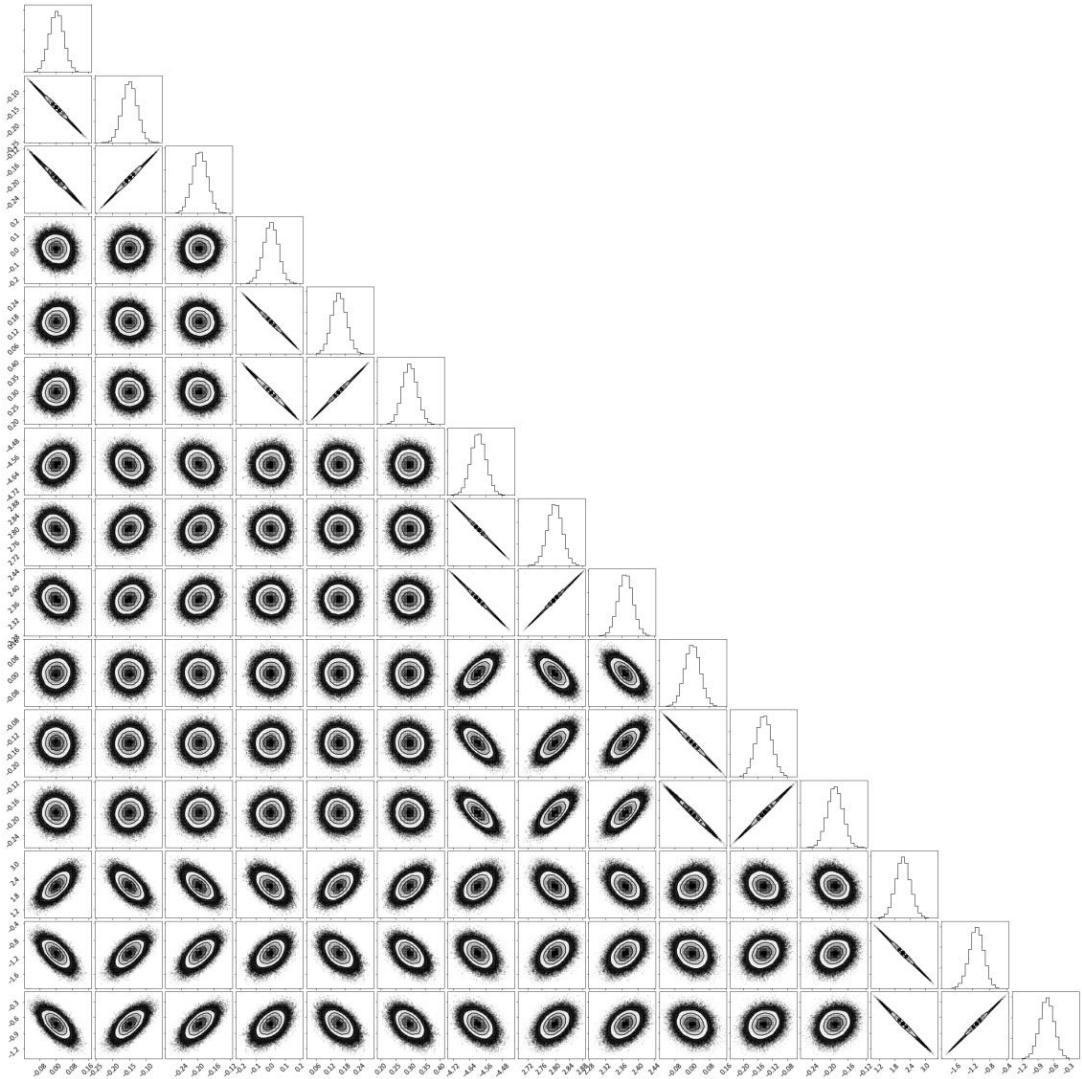

*Figure S4. 15 latent parameter distribution from the MCMC chain*

## V. SIMS and TR-PL results of the GaAs solar cell:

This section describes the experimental details of SIMS and TR-PL measurements of GaAs solar cells grown at different temperatures. The extracted doping concentration and bulk lifetimes serve as independent validation to the Bayesian network inference results shown in Figure 4 in the main body of the manuscript.

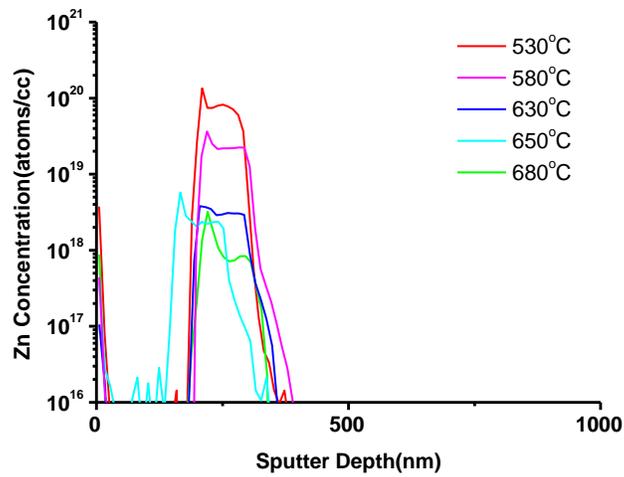

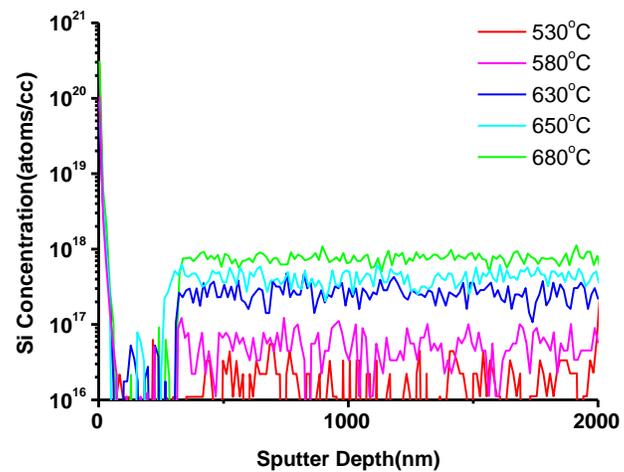

*Figure S5. Si and Zn dopant distribution in the GaAs solar cell under different growth temperatures.*

Figure S5 shows the zinc and silicon distributions in the emitter and bulk, respectively, as a function of growth temperature.

Table S1 shows the extrapolated bulk lifetime ($\tau$) from the *n*-doped InGaP/GaAs/InGaP heterostructure with two thicknesses (500 nm and 1000 nm) from the TR-PL measurement. The doping level and layer thickness for InGaP layers are identical to the back-surface field layer used in the GaAs solar cell structure. To extract the bulk lifetime of the n-doped GaAs, we numerically solve this differential equation for the excess carrier concentration as a function of distance to the front surface and time ($\Delta n(x,t)$). :

$$\frac{\partial \Delta n}{\partial t} = G + D \frac{\partial^2 \Delta n}{\partial x^2} - R \qquad [S1]$$

$G$ is the generation rate, $D$ is the diffusion coefficient, and $R$ is the recombination rate in Equation S1. We set the initial condition: $\Delta n(x, 0) = 0$ with boundary conditions where $S$ is the *n*-doped InGaP/GaAs surface recombination velocity and $d$ is the sample thickness:

$$\frac{\partial \Delta n}{\partial x}(x = 0, t) = \frac{S}{D} * \Delta n(x = 0, t); \quad \frac{\partial \Delta n}{\partial x}(x = d, t) = -\frac{S}{D} * \Delta n(x = d, t) \qquad [S2]$$

Hereafter, effective bulk lifetime is extracted by fitting the following equation.

$$\frac{1}{\tau_{eff}} = \frac{1}{\tau_{bulk}} + \frac{2S}{d} \qquad [S3]$$

*Table S1. Effective bulk lifetime of n-doped GaAs grown at different temperatures.*

| Growth temperature [°C] | $\tau$ [ns] |
|---|---|
| 530 | 1.8 |
| 580 | 12.5 |
| 630 | 29.3 |
| 650 | 11.1 |
| 680 | 0.6 |

## VI. Constraints for the Bayesian network prior:

Table S2 describes the constraints for the Bayesian network prior. Instead of directly setting constraints on *(a,b,c)*, we set hard constraints for the material properties that are computed from them in Equation [2] in the manuscript. The doping concentration, $\tau$ and *SRV* should have physically meaningful values. If they are not within those ranges shown in Table S2, the posterior probability will be set to negative infinity in the Bayesian network.

*Table S2. effective bulk lifetime of n-doped GaAs grown at different temperatures.*

| Parameter | Range |
| --- | --- |
| Zn doping concentration [cm$^{-3}$] | $1\times10^{16}$~$1\times10^{20}$ |
| Si doping concentration [cm$^{-3}$] | $1\times10^{16}$~$1\times10^{20}$ |
| Bulk lifetime [s] | $1\times10^{-10}$~$1\times10^{-6}$ |
| Front SRV [cm/s] | $1\times10^{2}$~$1\times10^{8}$ |
| Back SRV [cm/s] | $1\times10^{2}$~$1\times10^{8}$ |

## VII. Comparison between Bayesian network with black-box regression and Bayesian network with the Arrhenius prior:

Figure S6 shows the comparison of inferred material properties between the Bayesian network with black-box regression and the Bayesian network with the Arrhenius parametrization. We replace the physics-based Arrhenius parametrization shown in Equation 2 in the main text with a kernel ridge regression (KRR) model[6] using a radial basis function[6]. To test the impact of hyperparameters in the Bayesian network with KRR in the first layer, we vary the $\gamma$ value while keeping other parameters fixed. It can be shown that when $\gamma = 10$, the prediction from the Arrhenius parametrization and KRR are similar. As $\gamma$ decreases, the predictions from KRR for the bulk lifetime and the Rear *SRV* deviate from the predictions using the Arrhenius parametrization.

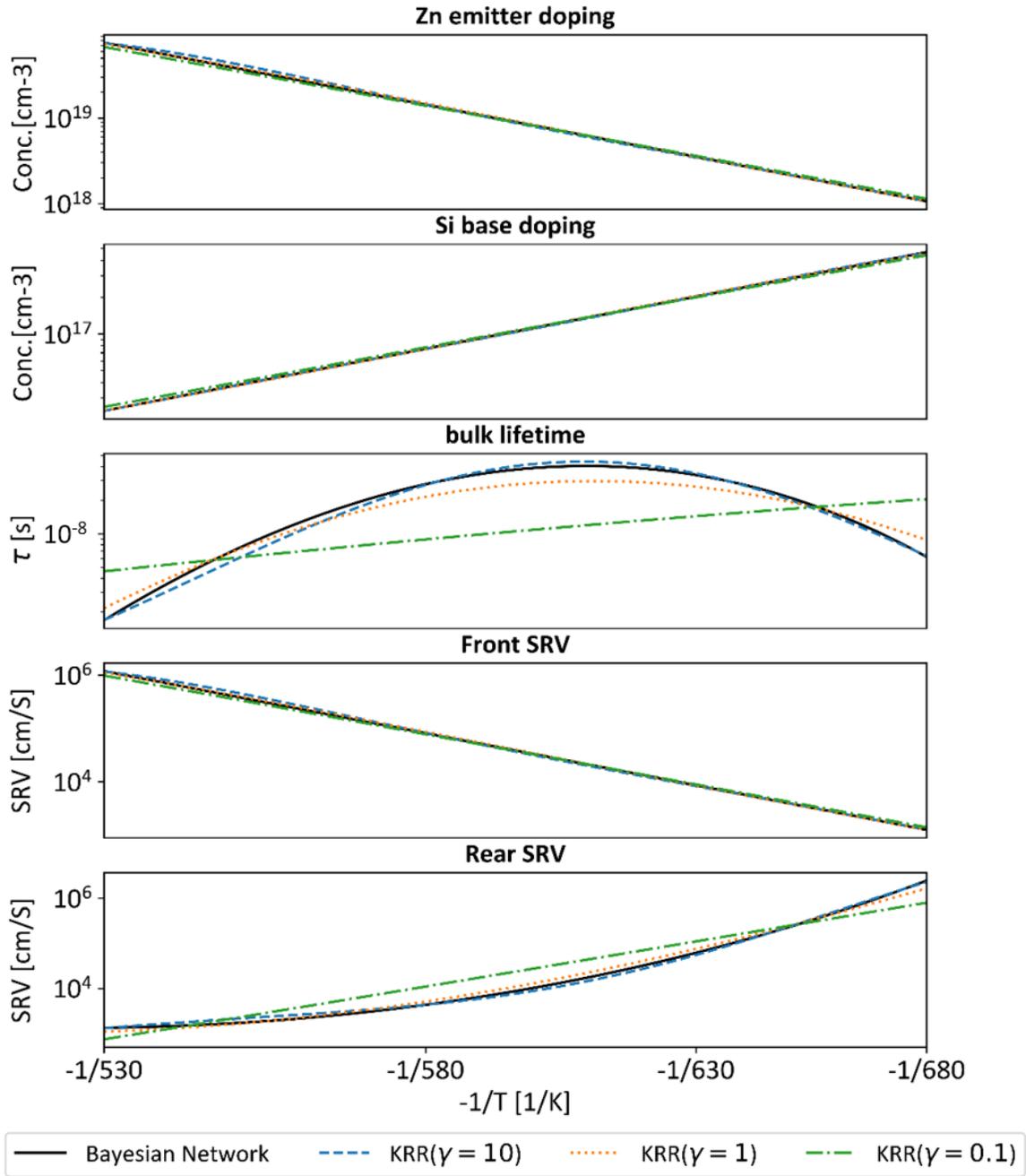

*Figure S6. inferred material properties between the Bayesian network with Kernel Ridge Regression (KRR) and the Bayesian network with the Arrhenius parametrization. The γ value of KRR is varied to show impact of hyperparameters on model* performance